\documentclass[12pt,a4paper]{article}
\usepackage[utf8]{inputenc}
\usepackage[T1]{fontenc}
\usepackage{amsmath,amssymb,bbm}
\usepackage{graphicx}

\begin{document}
\textwidth=135mm
 \textheight=200mm
\begin{center}
{\bfseries
 Importance of repulsive interactions for the equation of state and other properties of strongly interacting matter
 \footnote{{\small Talk at the 32th Max-Born-Symposium and HECOLS workshop on "Three Days of Phase Transitions in Compact Stars, Heavy-Ion Collisions
 and Supernovae", Institute for Theoretical Physics, University of Wroc\l{}aw, Wroc\l{}aw, Poland, February 17--19, 2014.}}}
 \vskip 5mm V.~V.~Begun$^{a,b}$ \vskip 5mm
 {\small {\it $^a$ Jan Kochanowski University, Kielce, Poland }}\\
 {\small {\it $^b$ Bogolyubov Institute for Theoretical Physics, Kiev, Ukraine}} \\
\end{center}
\vskip 5mm \centerline{\bf Abstract}
 We illustrate the role of repulsive interactions in a hadron-resonance gas at freeze-out and in a gas of quark-gluon bags.
 Taking into account non-zero size of particles in hadron gas leads to a significant decrease and shift of the net-baryon density maximum.
 The transition point from baryon to meson dominated matter depends on the difference between baryon and meson radii.
 We also show that depending on the properties of the quark-gluon bags one may obtain any type of the phase transition from hadron gas to quark-gluon plasma: the first or second order, as well as four types of the crossover.
 \vskip 10mm

\section{\label{sec:Intro}Introduction}
The hadron-resonance gas (HRG) model has become a standard tool for
the analysis of heavy ion collisions. With only a few parameters the
HRG model allows us to describe numerous ratios of particle
multiplicities produced in the wide energy range from SIS, the AGS  and SPS to RHIC and
the LHC~\cite{Becattini:2005xt,Cleymans:2005xv,Petran:2013lja,Stachel:2013zma}.
The proper definition of the space-time geometry and hydrodynamic
flow at freeze-out allows to further extend the HRG model to describe
the transverse-momentum spectra and other
soft-hadronic observables
\cite{Broniowski:2001we,Broniowski:2002wp,Begun:2014rsa}.
Resonances introduced to the ideal HRG effectively take into
account attractive interactions. However, the nucleon-nucleon potential
includes both attraction at large distances and repulsion at small
distances.
Repulsive interactions can be included in HRG using the van der Waals
excluded volume of particles~\cite{Gorenstein:1981fa}.
One should keep in mind that point-like hadrons would always
become dominant phase at very high energy density due to the large
number of different types of hadrons. Only the presence of the excluded volume
effects ensures a phase transition from the gas of hadrons and
resonances to the quark-gluon plasma (QGP)~\cite{Cleymans:1992jz}.
\section{\label{sec:HRG}Gas of extended hadrons and resonances}
In Ref.~\cite{Begun:2012rf} we calculated the excluded volume
effects along the chemical freeze-out line obtained for the
central Pb+Pb (Au+Au) collisions registered by experiments at SIS,
the AGS, SPS, and RHIC~\cite{Becattini:2005xt,Cleymans:2005xv}.
If all particles have the same volume, then the Boltzmann
approximation gives the same suppression factor for all densities,
which cancels in the particle ratios. However, the absolute value,
for example, of the net-baryon density strongly depends on the proper
volume and collision energy.

\begin{figure}[hbt]
   \centering
   \includegraphics[width=0.49\textwidth,keepaspectratio=true]{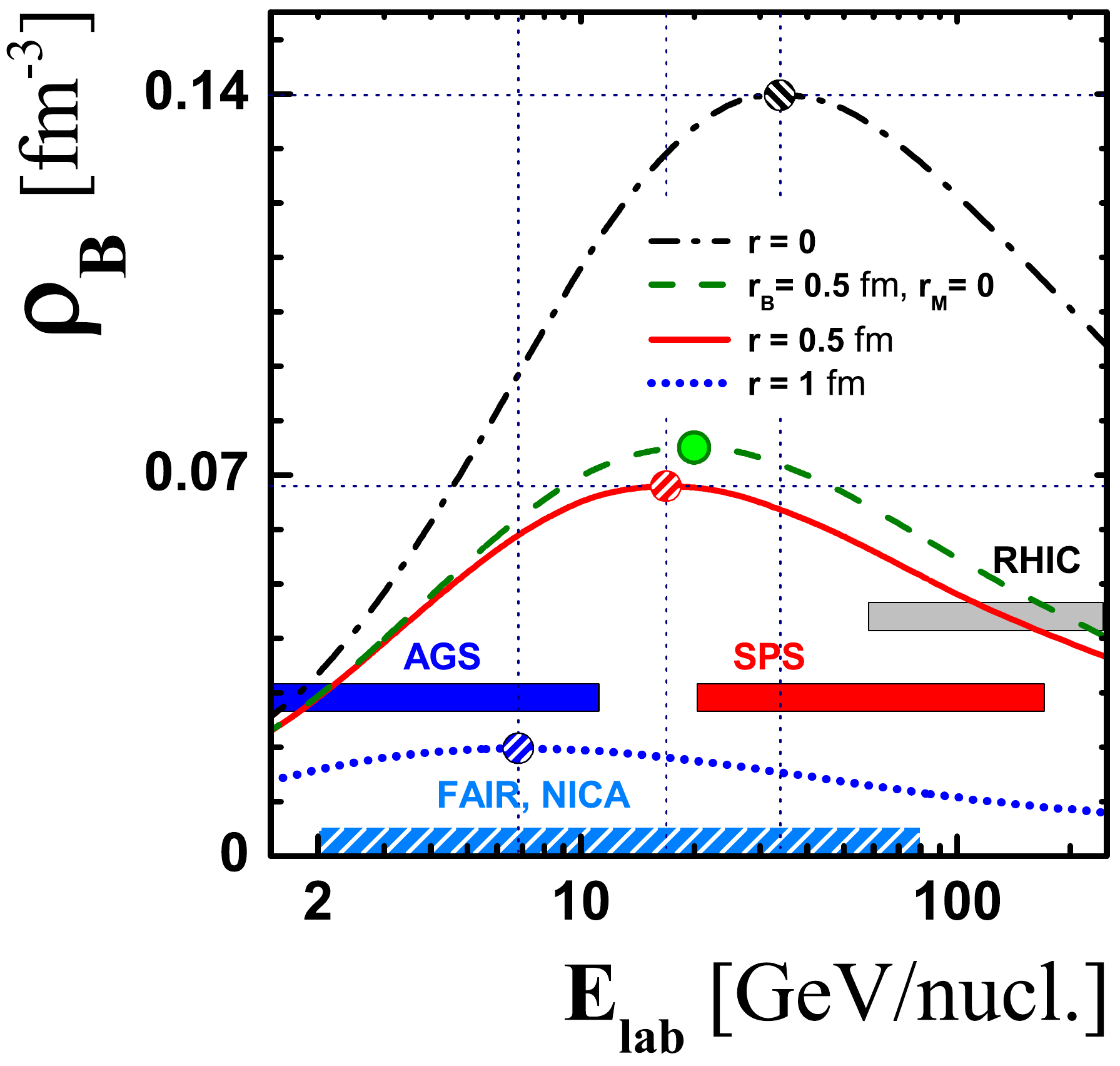}~
   \includegraphics[width=0.49\textwidth,keepaspectratio=true]{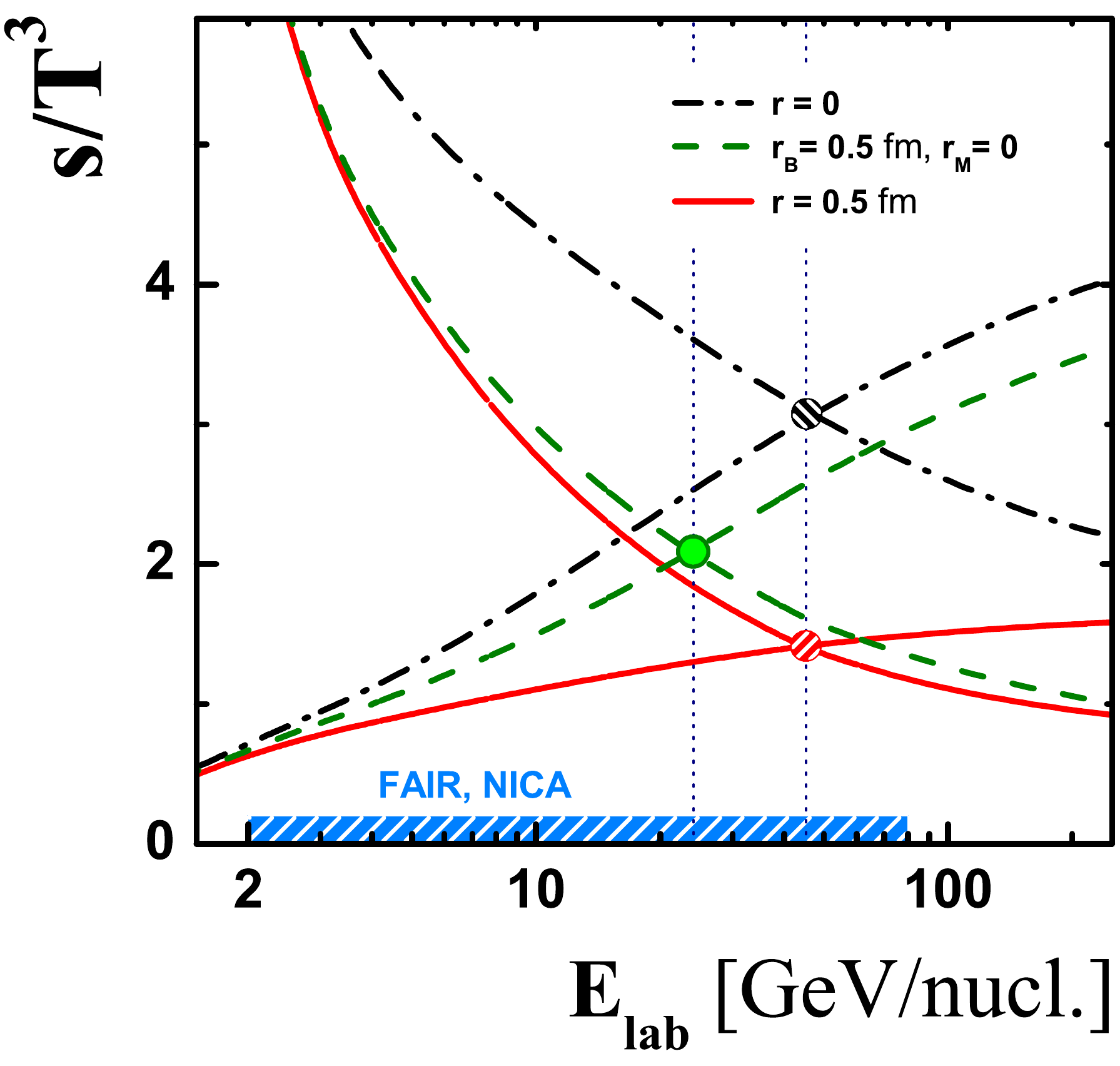}
   \caption{Left: The net-baryon density along the chemical
 freeze-out line~~\cite{Becattini:2005xt,Cleymans:2005xv}.
 Dashed-dotted line corresponds to the model of point-like particles with $r=0$,
 dashed line to $r_B=0.5$~fm and $r_M=0$,
 solid line to $r_B=r_M=r=0.5$~fm, and
 dotted line to $r_B=r_M=r=1$~fm.
 Right: The ratios of baryon and meson entropies to the temperature cubed,
 $s/T^3$, along the chemical freeze-out line.
 At small energies baryons dominate, while at large energies
 mesons take over.
 Dashed-dotted lines correspond to the model with $r=0$,
 solid lines to $r_B=r_M=r=0.5$~fm, and
 dashed lines to $r_B=0.5$~fm and $r_M=0$.
 The horizontal bands indicate the energy range of previous, current, and future experimental facilities for heavy-ion collisions.}
 \label{fig1}
\end{figure}

A moderate estimate of the particle radius by $r=0.5$~fm leads
to the suppression factor of about $0.75$ at the AGS and $0.5$
at the SPS, see~\cite{Begun:2012rf}. This non-monotonous
suppression with energy shifts the maximum of the net-baryon
density to lower energies, see the left panel of~Fig.~\ref{fig1}.
For the proper radii equal to $0.5$ and $1$~fm the density
maximum moves from $34$ to $17$ and $7$~AGeV, correspondingly.

The transition point from baryon to meson dominated matter is
sensitive only to the difference between baryon and meson
radii. It happens because at freeze-out the Boltzmann
approximation works well for entropy densities. Then, for equal
baryon and meson radii the corresponding entropies are
suppressed by the same factor and cross at the same point. For the
baryon radii $r_B=0.5$~fm and meson radii $r_M=0$, the
baryon/meson transition point moves from $46$ to $23$~AGeV,
see the right panel of~Fig.~\ref{fig1}.
\section{\label{sec:Cluster}Cluster Plasma}
One may extend the HRG model by adding the spectrum of bags filled
with non-interacting massless quarks and gluons to the usual HRG
spectrum. This is reasonable at high temperatures and densities,
where individual hadrons melt and merge, forming bigger states
with larger masses and volumes. In this way the bags may mimic the
effect of heavy resonances which have not been found yet. These
resonances/bags are not allowed to overlap, in the same way as usual
hadrons with the excluded volume. Interestingly, in such a system one can study the transition from HRG to
QGP analytically.

The possibility of phase transitions in the gas of quark-gluon
bags was demonstrated for the first time in Ref.~\cite{pt}.
Further studies allowed to obtain the 1st, 2nd, and higher order
transitions \cite{pt1,GGG,Zakout,Bugaev,Bessa}. A possibility of
no phase transitions was also pointed out in Ref.~\cite{pt1,GGG}. It was
suggested \cite{Ferroni} to model a smooth crossover transition by
the gas of quark-gluon bags. One can show that the type of the
transition from QGP to HRG is not sensitive to the presence of the
HRG, and depends only on the form of the mass-volume spectrum. In
Ref.~\cite{Begun:2009an} we summarized all known possibilities and
found even richer structure for the case of crossover.

\begin{figure}[hbt]
   \centering
   \includegraphics[width=0.49\textwidth,keepaspectratio=true]{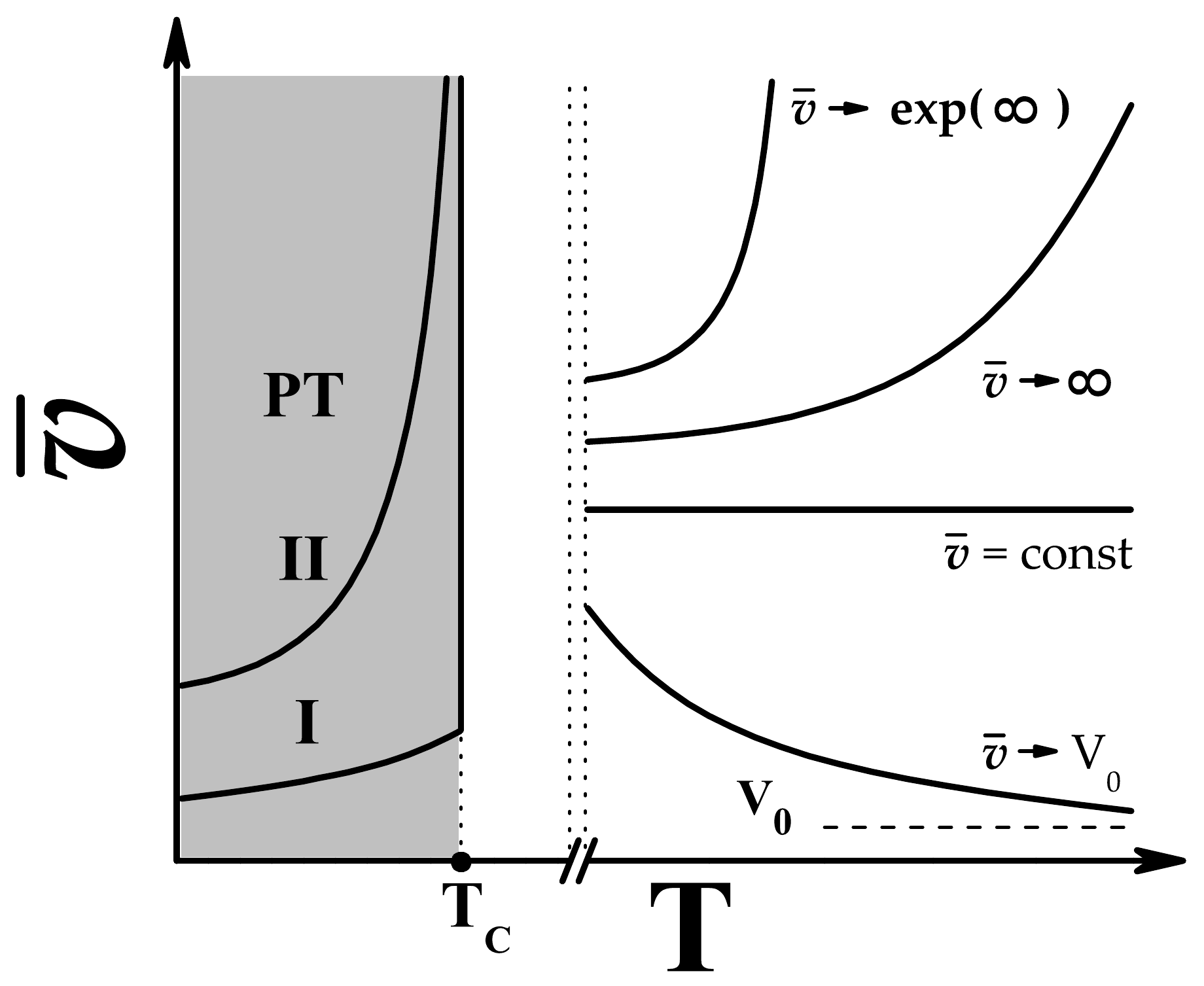}
     \caption{The schematic view of the possible temperature dependence of the average volume of the bag.
     The PT I and II labels denote the phase transition of the first and the second order, correspondingly.
     The right part of the figure shows the possible behavior of the bag after the crossover.}
   \label{fig2}
\end{figure}

The results are illustrated by the behavior of the average bag
size with increasing temperature, see Fig.~\ref{fig2}.
In the case of a phase transition the average bag size increases and
occupies the whole system. There are also two similar cases for
the case of crossover, see~\cite{Begun:2009an} for more details.
However, there are two more possibilities. In spite of the same
asymptotic behavior for the pressure and energy density, the
average bag size may be constant or even decrease to the minimal
value in thermodynamic limit.

\section{\label{sec:Cluster}Conclusions}
We have shown that the effect of taking into account the non-zero
proper volume of particles can be very strong. It leads to smaller
particle number and entropy densities and moves the position of
the net baryon density maximum to lower energies. The transition
point from baryon to meson domination is always situated at higher
energies than the density maximum. The transition point can move
to lower energies only for different meson and baryon radii in
the system. A more precise estimate of the baryon density
maximum and transition point can be important for the search on
the compressed baryonic matter at
GSI~\cite{Friman:2011zz,Galatyuk}.

The extended HRG model with quark-gluon bags can give any type of
the transition between HRG and QGP depending on the bag
properties. One may even get a cluster QGP, which is different
from the ideal QGP, despite the similar equation of
state.

\vskip 10mm \centerline{\bf Acknowledgment} This work was
supported by Polish National Science Center grant No.
DEC-2012/06/A/ST2/00390.

\end{document}